\documentclass{article}
\usepackage[nonatbib, preprint]{neurips_2024}
\usepackage{pdfpages}
\usepackage{float}
\usepackage{amsmath}
\usepackage{authblk}

\usepackage[utf8]{inputenc} 
\usepackage[T1]{fontenc}    
\usepackage{hyperref}       
\usepackage{url}            
\usepackage{booktabs}       
\usepackage{amsfonts}       
\usepackage{nicefrac}       
\usepackage{microtype}      
\usepackage{xcolor}         
\usepackage{graphicx}
\usepackage{gensymb}
\usepackage[capitalise]{cleveref}
\usepackage[font=small]{caption}
\usepackage{multirow}
\usepackage{footnote}
\usepackage[backend=biber,natbib=true,style=nature]{biblatex}
\usepackage{tikz}
\usetikzlibrary{shapes.geometric, arrows.meta, positioning, calc}

\urldef{\urlsfno}\url{https://catalog.ngc.nvidia.com/orgs/nvidia/teams/modulus/models/sfno\_73ch\_small}

\title{Data-driven Surface Solar Irradiance Estimation using Neural Operators at Global Scale}

\author[a, b, *]{A. Carpentieri}
\author[b]{J. Leinonen}
\author[b]{J. Adie}
\author[b]{B. Bonev}
\author[a]{D. Folini}
\author[b]{F. Hariri}

\affil[a]{\footnotesize{Institute for Atmospheric and Climate Science, ETH Zurich, 8092 Zurich, Switzerland}}
\affil[b]{\footnotesize{NVIDIA Corporation, 95051 Santa Clara, CA, USA}}
\affil[*]{\footnotesize{Corresponding author: acarpentieri@usys.ethz.ch}}

\begin{document}

\maketitle

\section*{Abstract}
Accurate surface solar irradiance (SSI) forecasting is essential for optimizing renewable energy systems, particularly in the context of long-term energy planning on a global scale. This paper presents a pioneering approach to solar radiation forecasting that leverages recent advancements in numerical weather prediction (NWP) and data-driven machine learning weather models. These advances facilitate long, stable rollouts and enable large ensemble forecasts, enhancing the reliability of predictions. Our flexible model utilizes variables forecast by these NWP and AI weather models to estimate 6-hourly SSI at global scale. Developed using NVIDIA Modulus, our model represents the first adaptive global framework capable of providing long-term SSI forecasts. Furthermore, it can be fine-tuned using satellite data, which significantly enhances its performance in the fine-tuned regions, while maintaining accuracy elsewhere. The improved accuracy of these forecasts has substantial implications for the integration of solar energy into power grids, enabling more efficient energy management and contributing to the global transition to renewable energy sources.

\begin{figure}[h]
    \centering 
    \includegraphics[width=\textwidth]{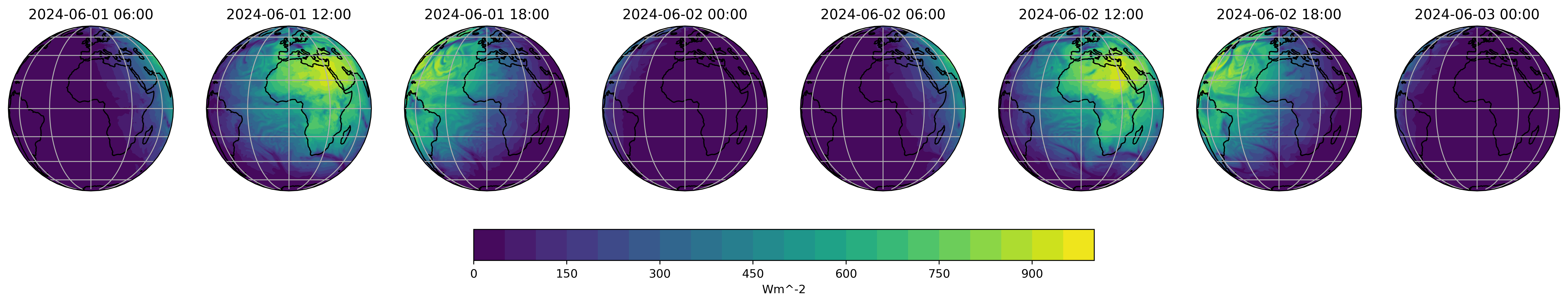} 
    \caption{\textbf{6-hourly averaged SSI forecasts over a 48-hour period}. SFNO \cite{SFNO, SFNO_checkpoint} acts as prognostic model forecasting multiple weather variables at 6-hourly lead times. Our diagnostic model is then applied to retrieve the accumulated 6-hour SSI. The proposed methodology is generic and can be applied to other weather forecasting models.} 
    \label{fig:ssrd_prediction} 
\end{figure}

\section{Introduction}
Surface solar irradiance (SSI) estimation and forecasting plays a crucial role in various applications, including renewable energy forecasting, agricultural planning, and climate studies. As the global push for sustainable energy sources intensifies, accurate prediction of solar irradiance becomes increasingly important for efficient integration of solar power into electricity grids \cite{bamisile2022, ANTONANZAS2016, NGUYEN2022}. 

Radiative transfer models have traditionally been used for SSI estimation, accounting for complex atmospheric interactions. For example, Urraca et al. \cite{URRACA2018} evaluated global horizontal irradiance - synonym for SSI - estimates from ERA5 and COSMO-REA6 against satellite-based radiative transfer models, emphasizing the need for accurate atmospheric data, including aerosol and cloud properties, to improve SSI estimates. The combination of satellite data with radiative transfer models has significantly enhanced their capabilities \cite{NGUYEN2022, Huang2019}. Alongside advancements in traditional radiative transfer models \cite{ukkonen2024twelve}, statistical and AI models have also been developed to further enhance the adaptability and speed of solar irradiance estimation approaches \cite{schuurman2024, LU2023, Gurel2023}.

The output of radiative transfer models (i.e. SSI fields) have been widely used as input to statistical and AI-based forecasting models \cite{PALETTA2023}. In fact, recent advancements in solar irradiance prediction have seen a shift towards AI-based models due to their ability to capture complex, non-linear relationships in meteorological data \cite{Papachristopoulou2024}, outperforming previous methods (i.e. optical flow models) in short-time horizons \cite{CARPENTIERI2023, CARPENTIERI2025}. 

While most current SSI forecasting models rely on satellite data and ground observations for accurate short-term forecasts, there is a growing need for SSI forecasting models capable of long-range predictions, commonly associated to lead times longer than few hours. In fact, by using satellite-based data as input, the models commonly forecast up to 6-hours in the future \cite{YANG2022743}. On the other hand, NWP and global AI weather models can produce forecast up to multiple days.

In \cite{molinaro2024}, a global AI weather model was trained to forecast SSI, outperforming the ECMWF (European Centre for Medium-Range Weather Forecasts) HRES predictions in forecasting energy-related variables such as wind and SSI. However, training and validating global weather models is expensive and not easily scalable to cover different variables. We propose a model that addresses this gap with the following novel contributions:

\begin{itemize}
    \item \textbf{Independence from satellite and ground observations}: Unlike many existing models, our approach does not rely on real-time satellite or ground-based measurements, enabling long-range forecasts by depending only on atmospheric variables commonly forecasted by NWP and AI-based weather models.
    \item \textbf{Integration with Advanced Weather Models}: Seamlessly incorporating numerical weather prediction (NWP) and AI weather models, leveraging advancements in weather prediction without needing to retrain or develop new weather models.
    \item \textbf{Enhanced Accuracy with Fine-Tuning}: Utilizing ERA5 pre-training and fine-tuning on satellite-derived data to significantly improve accuracy.
\end{itemize}

Our model delivers global, 6-hour average irradiance estimations at a high spatial resolution (0.25 degrees), supporting strategic energy planning and grid management. By combining physical models, satellite data, and AI techniques, we provide a comprehensive solution for global surface solar irradiance estimation, enhancing renewable energy forecasting and integration.

This paper is organized as follows: Section 2 provides a detailed description of the problem formulation, the data utilized, and the methods developed in our study. Section 3 presents the results obtained from our approach. In Section 4, we discuss the advantages and limitations of our approach and highlight potential opportunities for future research.

\section{Data and Methods}
The problem addressed in this work is formulated as follows: given the global atmospheric state \( x_{t_i} \) and additional context variables \( c_{t_i} \) at time \( t_i \), our goal is to predict the accumulated SSI global field \( y_{t_i} \) over the period \( \left[ t_i - \Delta, t_i \right] \) as described in \cref{eq:problem1} and \cref{eq:problem2}. 
\begin{equation} \label{eq:problem1}
y_{t_i} = f(x_{t_i}, c_{t_i})
\end{equation}
\begin{equation} \label{eq:problem2}
y_{t_i} = \frac{1}{\Delta} \int_{t_i - \Delta}^{t_i} \text{s}_t \, dt.
\end{equation}
where $f$ is the function we want to simulate, $s_t$ is the instantaneous SSI at time $t$ and $\Delta$ is the target aggregation time period, which is set to six hours to cover the most common weather forecasts frequency \cite{SFNO, lang2024aifs}. 
Additionally, by substituting \( x_{t_i} \) with the atmospheric state forecasted by a weather forecasting model, we can generate a global SSI forecast. This close link with  weather forecast models is advantageous, as it potentially allows to rapidly leverage progress in weather models for SSI forecast, in contrast to many existing SSI forecasting approaches that rely in part or entirely on data sources like satellite or ground based SSI observations. 

\subsection{ERA5} \label{sec:ERA5}
As input data for the model training, we used the ERA5 reanalysis dataset \cite{ERA5}, which is the fifth generation of ECMWF atmospheric reanalyses of the global climate. ERA5 provides hourly estimates of a large number of atmospheric, land, and oceanic climate variables, covering the Earth on a $0.25\degree$ grid and resolving the atmosphere using 137 vertical levels from the surface up to a height of 80 km.

ERA5 data are produced using ECMWF's Integrated Forecast System (IFS) Cy41r2, with 4D-Var data assimilation and 137 hybrid sigma/pressure (model) levels in the vertical reaching up to the top level at 0.01 hPa. Atmospheric data are available on these levels and they are also interpolated to 37 pressure, 16 potential temperature and 1 potential vorticity level(s).

Our study focuses on SSI, represented by the ERA5 Surface Solar Radiation Downwards (SSRD) variable. SSRD represents the amount of solar radiation reaching the Earth's surface, expressed in $\text{Jm}^{-2}$. It is accumulated over 1 hour and is a crucial parameter for various applications, including solar energy assessment, agricultural modeling, and climate studies. For simplicity, we transform the SSRD [$\text{Jm}^{-2}$] values to SSI [$\text{Wm}^{-2}$] estimates and use only SSI in the following.

The SSI in ERA5 is derived from a complex radiative transfer model that takes into account various atmospheric constituents, including clouds, aerosols, and gases. The accuracy of SSI estimates in ERA5 has been evaluated in several studies, such as \cite{URRACA2018}, showing comparable performance to satellite-based products on daily-aggregated data when compared to in situ surface based observations of SSI.

The ERA5 SSI data are hourly accumulations whereas our dataset uses 6-hourly values. We averaged six consecutive 1-hour values to produce a 6-hourly value for SSI. 

\subsection{SARAH3}
The SARAH-3 dataset, short for Surface Solar Radiation Data Set - Heliosat, Version 3 \cite{sarah3}, is an advanced satellite-based dataset providing detailed information on SSI. Developed by the Climate Monitoring Satellite Application Facility (CM SAF) of EUMETSAT, SARAH-3 employs the cloud index method derived from the original Heliosat-1 algorithm \cite{CANO1986}. This method calculates the effective cloud albedo (CAL) using the ratio between measured broadband reflectance, clear-sky reflectance, and maximum reflectance. The effective cloud albedo is then converted to a cloud index serving as input to a radiative transfer look-up table to retrieve the SSI. 

Instantaneous SSI from SARAH-3 is published with a two-week delay at a frequency of 30 minutes. The dataset spans several decades, providing consistent and continuous solar radiation data from 1983 onwards at 0.05$^\circ$. It is centered at (0$^\circ$N, 0$^\circ$E) and covers 65$^\circ$W to 65$^\circ$E and 65$^\circ$S to 65$^\circ$N. The validation of SARAH3 against ground-based measurements from the Baseline Surface Radiation Network (BSRN) \cite{sarah3} confirms its high accuracy and reliability, making it a trusted source for solar radiation data. 

The dataset comes on an equiangular grid at 0.05$^\circ$ resolution. Given that the input variables (from ERA5) for our estimation model come at $0.25^\circ$, we downsampled the SARAH3 dataset to match the ERA5 grid. For the spatial downsampling, we made use of a 
convolutional kernel. The kernel takes into consideration the different grids by weighting less the edge $0.05^\circ$ pixels that fall in multiple $0.25^\circ$ pixels so that the sum of radiation flux is matched in the downsampled dataset. In the Supplementary Material, we present our validation results and show that the downsampled SARAH3 dataset shows higher accuracy compared to the original version on most of the BSRN locations at 6-hour resolution.

\subsection{BSRN}
The Baseline Surface Radiation Network (BSRN) is a global network of high-quality surface radiation monitoring stations implemented by the World Climate Research Programme (WCRP) in 1992 \cite{BSRN}. The network aims to provide continuous, long-term, and frequently sampled measurements (data sampling frequency on the order of one minute) of surface radiation with the highest possible accuracy. BSRN stations are equipped with precision instruments, including pyranometers, to measure various components of solar radiation such as global horizontal irradiance (SSI), diffuse irradiance, and direct normal irradiance. These measurements are conducted with strict adherence to standardized protocols for instrument calibration, data acquisition, and quality control procedures to ensure data consistency and reliability across the network. The BSRN data are widely used for climate research, satellite validation, and the study of long-term trends in surface radiation.

\subsection{Solar Irradiance Diagnostic Model}
The purpose of our diagnostic model is to provide a surface solar irradiance forecast exploiting the recent progress in numerical and AI-based weather predictions. The pool of available ERA5 variables is thus defined by the variables provided by the aforementioned forecast models. As reference, we take the 73 output variables defined in \cite{SFNO}, which comprises 7 surface variables and 5 variables at 13 pressure levels. Differently from \cite{SFNO}, the relative humidity $r$ is replaced with specific humidity $q$, compatible with the new SFNO checkpoint \cite{SFNO_checkpoint}.

Accurate surface solar irradiance estimation requires consideration of several variables at multiple atmospheric layers, as well as surface and column integrated variables. Our specific choice of variables is motivated by the following considerations. While lower tropospheric levels (1000-850 hPa) are crucial due to their high water vapor and aerosol content \cite{Kosmopoulos2018, Gurel2023}, mid and upper tropospheric levels also play significant roles in capturing cloud information and overall atmospheric structure. 
In fact, while lower levels contain most of the atmospheric water content, mid-levels (800-500 hPa) still hold a significant amount of water vapour \cite{Reynolds1975}. On the other hand, upper levels (300-200 hPa) provide information on high-level clouds and cirrus, which can attenuate incoming solar radiation \cite{Gurel2023}. Given the importance of the different levels, we make use of more pressure levels closer to the surface (1000, 925, 850 hPa) and fewer levels at higher altitudes (700, 500, 300, 50 hPa) but still covering the full range of available pressure levels. For the different pressure levels, we retrieved the geopotential height $z$, the temperature $t$ and the specific humidity $q$. It is important to note that temperature combined with specific humidity indicates the dew point and thus the potential for condensation and cloud formation. Moreover, we also make use of surface variables, specificallysurface temperature ($T2m$) and surface pressure ($Sp$), as well as of total column of water vapor ($tcwv$). The input variables are summarized in \Cref{tab:variables}.

\begin{table}[ht]
\centering
\resizebox{.6\textwidth}{!}{%
\begin{tabular}{l|c}
\toprule
\textbf{Variable} & \textbf{Levels} \\
\midrule
\textbf{Geopotential Height $z$} & 50, 300, 500, 700, 850, 925, 1000 \\
\midrule
\textbf{Temperature $t$} & 50, 300, 500, 700, 850, 925, 1000, 2 m \\
\midrule
\textbf{Specific Humidity $q$} & 50, 300, 500, 700, 850, 925, 1000 \\
\midrule
\textbf{Total column of water vapor} & Vertically integrated \\
\midrule
\textbf{Pressure} & Surface \\
\bottomrule
\end{tabular}%
}

\caption{\textbf{ERA5 input variables.} The input variable selection took in consideration the set of forecasted variables of SFNO \cite{SFNO}. $z$, $t$ and $q$ are sampled at the most relevant pressure levels (shown in hPa). The estimation model is also conditioned on orography, positional embeddings and solar zenith angle by concatenating a global digital elevation model, sinusoidal embeddings of latitude and longitude and the cosine of the solar zenith angle computed for every pixel.}
\label{tab:variables}
\end{table}

Based on the precipitation estimation model developed in FourCastNet \cite{pathak2022}, our model architecture for SSI estimation is also based on Adaptive Fourier Neural Operators (AFNO) \cite{guibas2021} as shown in \Cref{fig:architectures}. However, with respect to the AFNO architecture presented in \cite{pathak2022}, we modify the patch embedding strategy and remove the learnable positional embeddings. The new patch embedding method implemented in our SSI estimation model uses convolutional layers, which wrap around the longitudinal axis, reflecting the spherical shape of the Earth \cite{siddiqui2024}. This symmetrical patch embedding is done to reduce possible artifacts the AFNO can produce at the corners of the image and at the patch edges. Secondly, we remove the learnable positional embedding in order to permit the model to run on arbitrarily large images and replace it instead with sinusoidal embeddings of the latitude and longitude coordinate grids.

In our AFNO architecture (see \Cref{fig:architectures}), data flow begins with the symmetrical patch embedder. This component maps each \(4 \times 4\)-pixel patch into a single latent vector. The input atmospheric state \(x \in \mathbb{R}^{n \times m \times c}\) is initially symmetrically padded to \(\tilde{x} \in \mathbb{R}^{(n + p_n) \times (m + p_m) \times c}\). It is then transformed into a latent state \(v \in \mathbb{R}^{\frac{(n + p_n)}{4} \times \frac{(m + p_m)}{4} \times 256}\). Here, \(n\), \(m\), and \(c\) denote the longitude, latitude, and channel dimensions of the atmospheric state, respectively. For our specific use case at ERA5 resolution, these dimensions are 1440, 721, and 31, respectively. \(p_n\) and \(p_m\) represent the padding sizes for the longitude and latitude dimensions. The embedded representation $v$ is processed by 12 AFNO blocks and ultimately projected to the target shape. The AFNO blocks are as described in \cite{pathak2022}, as shown in \Cref{fig:architectures}.

\subsubsection*{Training}
We train our diagnostic model on 37 years of 1-hourly ERA5 data (1980--2016) and validated on one year (2017). The input variables are shown in \Cref{tab:variables}, while the output variable is the shortwave surface downwelling radiation product aggregated to 6-hour temporal resolution as described in \Cref{sec:ERA5}. The loss function is a weighted $L_2$ loss as defined in \cref{eq:weighted_L2_loss}.

\begin{equation}
L_2^{\text{weighted}} = \frac{1}{\sum_{i=1}^N w(\phi_i)} \sum_{i=1}^N w(\phi_i) (y_i - obs_i)^2
\label{eq:weighted_L2_loss}
\end{equation}

\begin{equation}
w(\phi_i) = \text{cos}(\phi_i)
\label{eq:weight_func}
\end{equation}

where $y_i$ and $obs_i$ are the prediction and ERA5 observation SSI values for the $i-th$ grid box. $N$ is the number of grid boxes in the SSI field. The weights $w(\phi_i)$ depend on the latitude $\phi_i$ as defined in \cref{eq:weight_func}, where $\phi_i$ ranges from $-90^\circ\text{N}$ to $90^\circ\text{N}$.

The training is governed by a a cosine annealing scheduler starting from a learning rate of $10^-3$ and run for 30 epochs on 32 NVIDIA A100 GPUs with batch size set to 128. We use the Adam optimizer \cite{kingma2014adam}. The full model training took approximately 512 GPU hours.

\subsubsection*{Fine-tuning}
To further enhance the accuracy of our SSI diagnostic model, we fine-tuned it on the satellite-based SARAH3 dataset. The main challenge of this was posed by the comparatively small domain of the dataset, which covers approximately $1/4$ of the full globe. The finetuning is motivated by the higher accuracy provided by the SARAH3 SSI fields compared to ERA5 \cite{URRACA2018}.

Once the model is trained on the ERA5 SSI product, we fine-tune the AFNO-based diagnostic model on the SARAH3 $0.25^\circ$ dataset (2000--2016) for 10 epochs by using a masked $L_2$ loss, which computes the residuals only on the SARAH3 domain. The fine-tuning is accomplished by starting training from the $\text{AFNO}_{\text{ERA5}}$ pre-trained weights and with a low learning rate of $5\times10^{-5}$ and with a cosine annealing scheduler. The finetuning is again run on 32 NVIDIA A100 GPUs with batch size set to 128, taking approximately 80 GPU hours to complete.

\begin{figure}[htbp]
    \centering
        \includegraphics[clip, trim=0cm 19cm 1cm 2cm, width=1.00\textwidth]{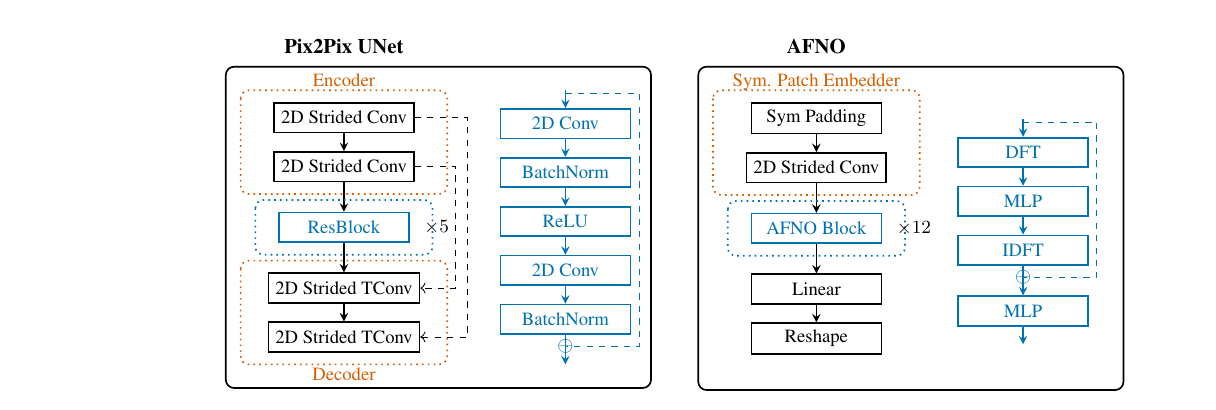}
    \caption{\textbf{On the left}: the Pix2Pix UNet architecture with residual blocks (ResBlock) showed in blue. The encoder and decoder are composed by strided convolutions (2D Conv) and strided transposed convolutions (2D TConv). The first encodes the weather forecasted fields to the latent space while the second decodes the latent prediction into a SSI field. \textbf{On the right}: the AFNO architecture with AFNO blocks showed in blue. The encoding in the latent space is obtained by symmetrically padding the input fields and successively applying a strided convolution. Then, the latent fields are processed through AFNO blocks and mapped to SSI by a linear layer.}
    \label{fig:architectures}
\end{figure}


To differentiate the AFNO-based models trained on different datasets, we name $\text{AFNO}_{\text{ERA5}}$ and $\text{AFNO}_{\text{SARAH3}}$ the AFNO-based models trained on the ERA5 and SARAH3 SSI datasets. $\text{AFNO}_{\text{f}}$ refers to $\text{AFNO}_{\text{ERA5}}$ finetuned on SARAH3.

\section{Results}
To test the proposed global SSI diagnostic model, we compare it against two benchmarks models: a convolutional U-Net based on the Pix2Pix architecture \cite{pix2pix} shown in \Cref{fig:architectures}, and a point-wise estimation model built on a stack of multi-layer perceptron (MLP) blocks. Both benchmark models were trained on the same data and used the same training schedule as our AFNO architecture. The Pix2Pix UNet architecture consists of two convolutional encoding blocks, five ResNet layers and two convolutional decoding blocks. Each encoding block reduces the spatial dimension by a factor of 2. The U-Net latent space reflects the AFNO model embedding strategy for a fair comparison: every latent pixel correspond to a $4\times4$ input patch and is represented by a vector in $R^{256}$. On the other hand, the MLP-based model is composed of 5 MLP layers: the first layer embeds the input into the latent space and the last layer remaps the latent vector to one single scalar representing the aggregated SSI for one single pixel. A single MLP layer is composed by two linear layers and one ReLU activation function in between.

For the model validation we run the diagnostic models on 2018 ERA5 data at 6-hourly intervals (00:00, 06:00, 12:00, 18:00 UTC), which are commonly used to initialize most weather forecasting models. Model performance is measured on ERA5, SARAH3 and BSRN 6-hourly aggregated SSI observations. The validation on ERA5 is performed to measure the accuracy of the different architectures in learning the relationship between the input features and SSRD. ERA5 is the least accurate estimate \cite{URRACA2018} but covers the full globe. SARAH3 provides more accurate irradiance estimates but it is limited to the EUMETSAT geostationary satellite domain. Finally, BSRN pyranometers provide direct measurements but only on few locations from which we can have only sparse metrics. Moreover, BSRN data are point measurements, and some differences compared to spatially aggregated pixel or gridded data are to be expected \cite{URRACA2018}.

\subsection{The need for large receptive fields} \label{sec:ERA5_validation}
\begin{figure}[h] 
    \centering 
    \includegraphics[width=\textwidth]{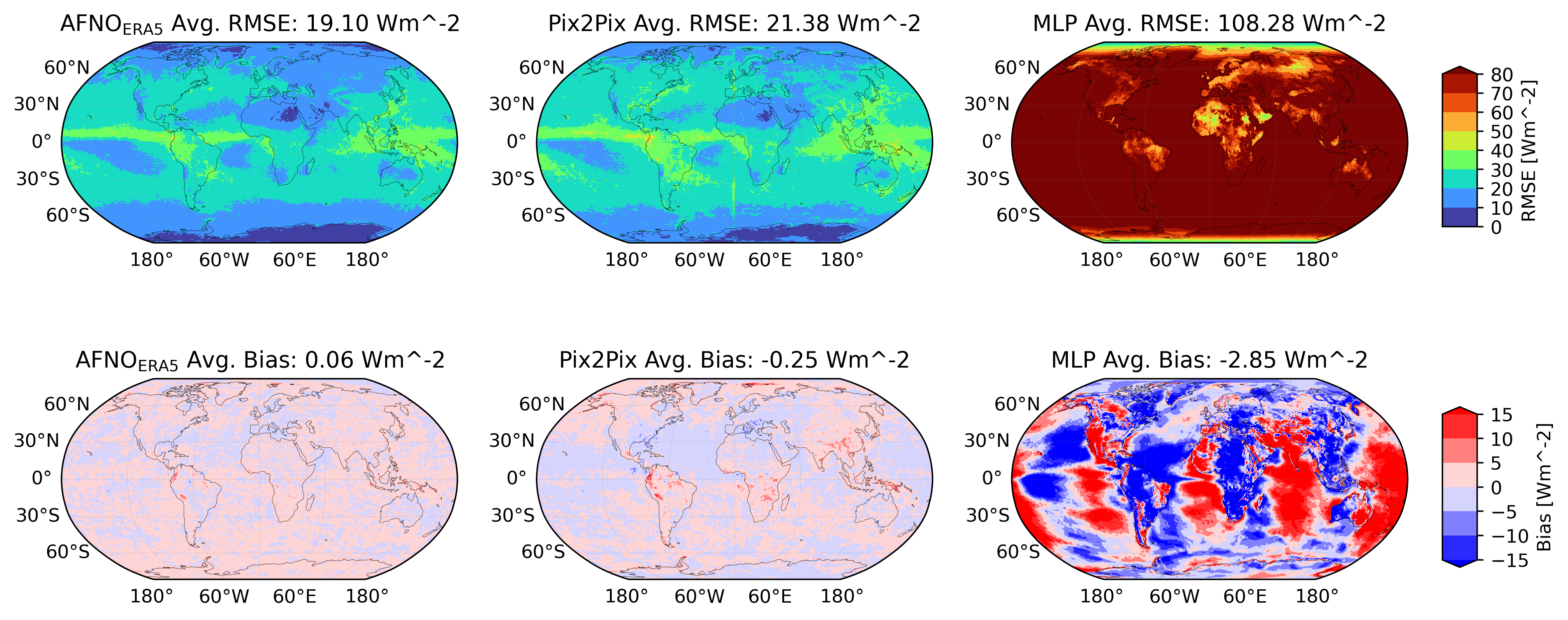} 
    \caption{\textbf{Evaluation vs.~ERA5 6-hour SSI}. Our model ($\text{AFNO}_{\text{ERA5}}$) is evaluated against ERA5 SSI product and compared against two benchmark models (Pix2Pix and MLP), likewise evaluated against the ERA5 SSI product. The validation is performed on 6-hour averaged fields of solar radiation for 2018.} 
    \label{fig:era5_metrics} 
\end{figure}

Our task involves the estimation of an aggregated variable over the past 6 hours. Differently from instantaneous estimation as done in \cite{schuurman2024}, we need to predict previous atmospheric states in order to integrate them and retrieve the aggregated output. Clouds are the main irradiance blockers, so our task can be naively translated into predicting the cloud locations in the previous 6 hours based on the current model forecast state. Consequently, the model receptive field - the image region affecting the prediction of a single pixel - is important for capturing such advective cloud dynamics. In \Cref{fig:era5_metrics}, the three models are ordered based on their receptive field size. The MLP model works on the single-pixel level, making it difficult for the model to capture the variation of solar radiation in time, especially because it captures the solar zenith angle and the atmospheric state only for one location and for one timestamp in time. On the other hand, the convolutional model (Pix2Pix) has a receptive field of $25\times25$ pixels -- $7\times7$ from the encoder block plus $18\times18$ from the 5 latent residual blocks with two convolutional layers each --, making it able to capture past atmospheric evolutions. The performance is further enhanced by the global receptive field of the AFNO model that perform global convolution by working in the Fourier Space. 

Even if $\text{AFNO}_{\text{ERA5}}$ and Pix2Pix models show similar error patterns (see \Cref{fig:era5_metrics}), $\text{AFNO}_{\text{ERA5}}$ outperforms Pix2Pix by $10.7\%$ and $76\%$ in terms of root mean squared error (RMSE) and bias, respectively.

\subsection{Fine-tuning evaluation}
Given the validation on daily aggregations performed in \cite{URRACA2018}, satellite-based solar irradiance estimates (i.e. SARAH) are more accurate than reanalysis datasets (i.e.~ERA5). In \Cref{tab:bsrn_era5_sarah3_comparison}, we confirm the results also for 6-hourly aggregation data. Moreover, we show that for such temporal frequencies, coarsening the resolution of the satellite-based dataset does not degrade the accuracy, but leads to slightly improved performance over most stations. 

\begin{figure}[ht] 
    \centering 
    \includegraphics[width=\textwidth]{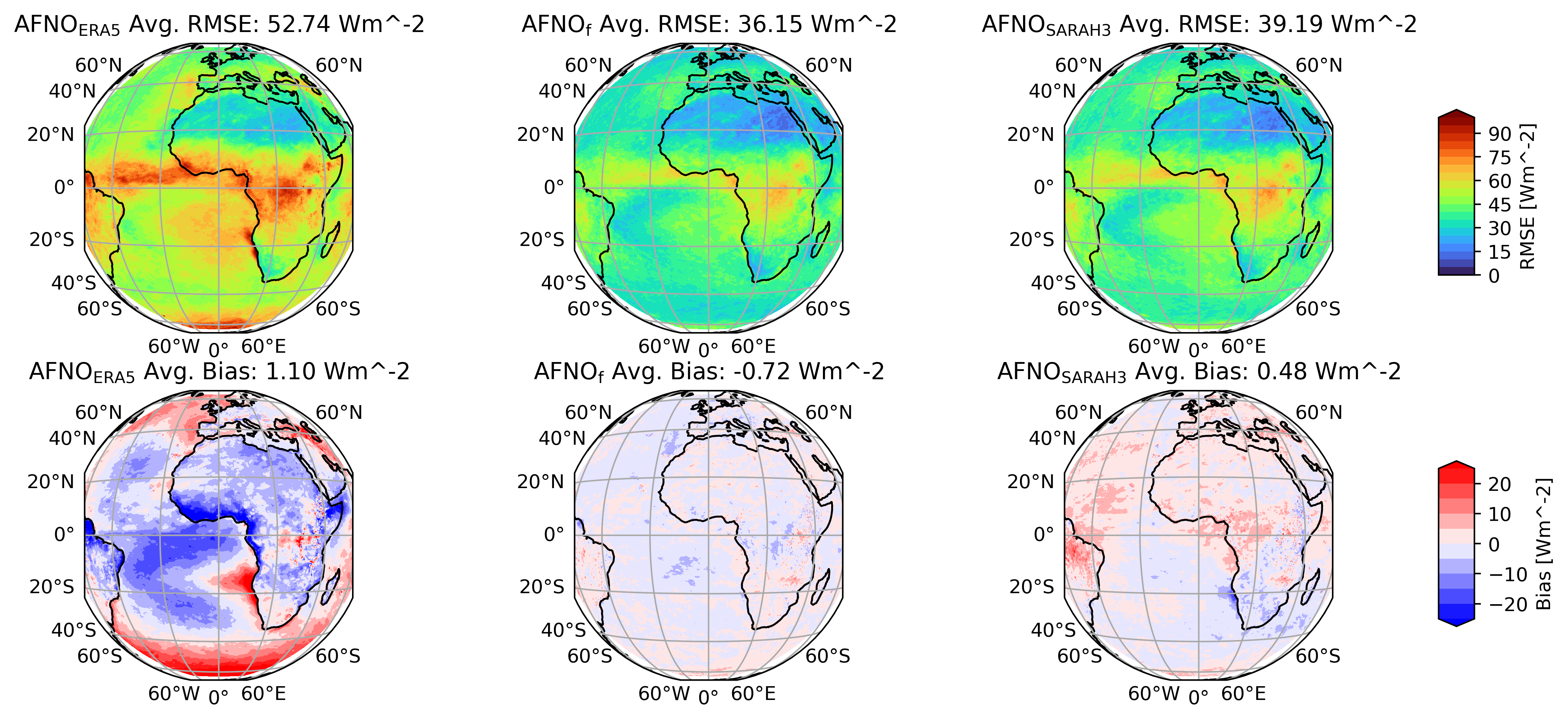} 
    \caption{\textbf{Evaluation vs SARAH3 6-hour SSI}. Three version of our AFNO-based model are validated and compared on SARAH3 6-hour averaged SSI fields at $0.25^\circ$ resolution for 2018. $\text{AFNO}_{\text{ERA5}}$ is the ERA5-trained version, the second model is the same model finetuned on 17 years of SARAH3 SSI fields ($\text{AFNO}_{\text{f}}$), while $\text{AFNO}_{\text{SARAH3}}$ has the same architecture but it is fully trained on SARAH3 without pretraining on ERA5 irradiance product.} 
    \label{fig:sarah3_metrics} 
\end{figure}

Motivated by the higher accuracy of satellite datasets, we train our AFNO-based estimation model on the SARAH3 dataset ($\text{AFNO}_{\text{SARAH3}}$). Furthermore, we fine-tune the ERA5-trained AFNO model on the same SARAH3 dataset ($\text{AFNO}_{\text{f}}$). In \Cref{fig:sarah3_metrics}, we show RMSE and bias results of the aforementioned models and $\text{AFNO}_{\text{ERA5}}$, which is trained only on ERA5. 

The higher RMSE shown by $\text{AFNO}_{\text{SARAH3}}$ over SARAH3 shown in \Cref{fig:sarah3_metrics} with respect to $\text{AFNO}_{\text{ERA5}}$ over ERA5 (see \Cref{fig:era5_metrics}) suggests that estimating SARAH3 SSI values from ERA5 inputs is a more difficult task than estimating ERA5 SSI from the ERA5 input data. This is probably due to inaccuracies present in the ERA5 dataset leading to difficulties in estimating direct observation variables. Meanwhile, estimating ERA5 SSI from ERA5 inputs is relatively easy as the diagnostic model only needs to learn to emulate the ERA5 scheme for SSI calculation. However, the main result of our analysis is the fact that fine-tuning the ERA5-trained model ($\text{AFNO}_{\text{ERA5}}$) on direct observation results in higher accuracy than directly training the model on SARAH3 observational data. In fact, $\text{AFNO}_{\text{f}}$ shows a $31.5\%$ and a $7.8\%$ RMSE improvement over $\text{AFNO}_{\text{ERA5}}$ and $\text{AFNO}_{\text{SARAH3}}$, respectively. 

\subsection{Ground evaluation}
The most accurate measurements of SSI are retrieved by pyranometers located at ground weather stations. 
In particular, BSRN provides the most accurate surface solar radiation measurements due to their meticulous quality control. To validate against the ground stations, we picked the pixel surrounding the station and compared it to the station observations. 

Differently from \Cref{tab:bsrn_era5_sarah3_comparison}, we validate the models on 2018 at 6-hour resolution, using 6-hour aggregated measurements at 00, 06, 12 and 18 UTC. Only BSRN stations with less than 10\% of missing data are kept for the evaluation.

As reference, we use the ERA5 and SARAH3 metrics measured on the ground station locations. As described in \Cref{sec:ERA5_validation}, AFNO demonstrates to be the best performing architecture. Comparing to ground observations, $\text{AFNO}_{\text{ERA5}}$ and Pix2Pix show comparable skill as ERA5. The slight improvement over the ERA5 dataset could be due to the fact that the model tend to predict smoother fields than the reference data (ERA5), thus reducing the error especially when ERA5 and ground observations largely disagree.

On stations located within the SARAH3 domain, SARAH3 shows higher accuracy ($27.6\%$ improvement over ERA5 in RMSE), as also demonstrated in \cite{URRACA2018} and \Cref{tab:bsrn_era5_sarah3_comparison}. In fact, the SARAH3-trained AFNO shows a slight improvement over the ERA5-trained version at the cost of a much worse accuracy outside the SARAH3 domain. On the other hand, the fine-tuned version not only demonstrates higher accuracy compared to SARAH3 (see \Cref{fig:sarah3_metrics}) but also a significant improvement on ground stations: $8.2\%$ lower RMSE compared to $\text{AFNO}_{\text{ERA5}}$. Moreover, $\text{AFNO}_{\text{f}}$ shows to be stable on locations outside the fine-tuning domain. 

\begin{table}[ht]
\centering
\resizebox{.6\textwidth}{!}{%
\begin{tabular}{l|l|c|c|c}
\toprule
\textbf{Metric} & \textbf{Model} & \textbf{All Stations} & \textbf{SARAH3 Stations} & \textbf{ROW Stations} \\
\midrule
\multirow{7}{*}{\textbf{RMSE}} & ERA5 & 52.60 & 52.40 & 52.72 \\
 & SARAH3 & & 37.95 & \\
 & $\text{AFNO}_{\text{ERA5}}$ & 50.81 & 50.90 & 50.76 \\
 & $\text{AFNO}_{\text{f}}$ & \textbf{48.68} & \textbf{46.72} & \textbf{49.84} \\
 & $\text{AFNO}_{\text{SARAH3}}$ & 79.75 & 49.22 & 97.78 \\
 & Pix2Pix & 50.85 & 51.72 & 50.33 \\
 & MLP & 107.13 & 108.33 & 106.43 \\
\midrule
\multirow{7}{*}{\textbf{MAE}} & ERA5 & 26.29 & 27.16 & 25.79 \\
 & SARAH3 & & 20.52 & \\
 & $\text{AFNO}_{\text{ERA5}}$ & \textbf{25.93} & 27.20 & \textbf{25.19} \\
 & $\text{AFNO}_{\text{f}}$ & 26.86 & \textbf{25.72} & 27.53 \\
 & $\text{AFNO}_{\text{SARAH3}}$ & 50.83 & 27.36 & 64.70 \\
 & Pix2Pix & 26.26 & 27.94 & 25.27 \\
 & MLP & 69.27 & 68.03 & 70.00 \\
\midrule
\multirow{7}{*}{\textbf{Bias}} & ERA5 & 1.61 & -5.45 & 5.78 \\
 & SARAH3 & & -2.36 & \\
 & $\text{AFNO}_{\text{ERA5}}$ & 0.79 & -5.96 & 4.78 \\
 & $\text{AFNO}_{\text{f}}$ & -5.35 & -3.89 & -6.21 \\
 & $\text{AFNO}_{\text{SARAH3}}$ & 29.96 & \textbf{-1.65} & 48.65 \\
 & Pix2Pix & \textbf{0.23} & -6.46 & \textbf{4.18} \\
 & MLP & -4.39 & -20.41 & 5.08 \\
\bottomrule
\end{tabular}%
}
\label{tab:performance_metrics}
\caption{\textbf{Estimation accuracy metrics on the BSRN locations}: RMSE, MAE and bias $[\text{Wm}^2]$ are presented for the estimation models and gridded datasets across all stations, SARAH3-domain stations, and rest-of-the-world (ROW) stations. The lowest (best) values among the models (excluding SARAH3 and ERA5) are highlighted in bold.} 

\end{table}

\section{Conclusion}
We present a novel model for global solar radiation estimation. By leveraging advancements in numerical weather prediction (NWP) and machine learning (ML) weather models, our model is able to provide stable medium-term SSI forecasts at a global scale. Moreover, our predictions are independent of real-time satellite and ground-based observations, which enables more reliable and extensive forecasts by depending only on the weather forecasting model output. By seamlessly integrating NWP and ML weather models, the framework enhances accuracy and reliability, thus enabling flexible deployment in various operational environments. Finally, we show how fine-tuning with satellite-derived data significantly improves the model's performance, especially in specific regions. 

Delivering 6-hourly aggregated SSI estimates on a global scale, our model supports strategic energy planning and grid management, potentially impacting the efficient integration of solar energy into power grids, aiding the global transition to renewable energy sources. As such, future research should focus on refining these techniques and expanding the model's capabilities to enhance renewable energy forecasting even further.

\section*{Code \& Data Availability}
The ERA5 dataset is available and freely accessible at: https://cds.climate.copernicus.eu/.
The Heliosat SARAH3 dataset is available at: https://wui.cmsaf.eu/.
The BSRN ground stations dataset is available at: https://bsrn.awi.de/.
For the model development, training and validation, we refer to the Modulus open-source repository: https://github.com/NVIDIA/modulus.

\printbibliography[title={References}]
\newpage
\clearpage

\renewcommand\thefigure{\thesection.\arabic{figure}}
\renewcommand\thetable{\thesection.\arabic{figure}} 
\setcounter{figure}{1}
\appendix
\section{Supplementary Material}
\subsection{Comparison of ERA5 and SARAH3 at BSRN Locations}
This section presents a detailed comparison of SSI data from ERA5 and SARAH3 against observations from the BSRN stations covered by the SARAH3 domain. We focus on evaluating the accuracy of these datasets using key metrics such as Root Mean Square Error (RMSE), Mean Absolute Error (MAE), and Bias. We compared only 6-hourly aggregated SSI measurements relevant for our application and compatible with most of weather forecasting models temporal resolution. The observations are retrieved for all the stations falling inside both domains and for the (2015, 2016) period.

The results indicate that SARAH3 generally exhibits superior performance compared to ERA5 across most BSRN locations as showed in \Cref{tab:bsrn_era5_sarah3_comparison}. Specifically, SARAH3 achieves lower RMSE values, suggesting enhanced predictive accuracy. The downsampled SARAH3 dataset at $0.25^\circ$ resolution often outperforms the original $0.05^\circ$ resolution, except at the \textsl{tam} location, where the original resolution shows better results.

In terms of MAE, SARAH3 consistently demonstrates lower values than ERA5, further confirming its higher accuracy in estimating solar irradiance. However, ERA5 tends to have smaller absolute bias values and SARAH3 maintains a consistent positive bias across several stations.

The comparison between different resolutions of SARAH3 reveals that the coarser $0.25^\circ$ resolution marginally improves upon the $0.05^\circ$ resolution in both RMSE and MAE at most sites. This suggests that the downsampled version effectively balances spatial detail with accuracy.

The performance of these datasets varies geographically. Notably, the \textsl{iza} station, located in Izaña, Canary Islands, exhibits significantly higher errors for both datasets. This can be attributed to its high altitude (2373 meters) and unique climatic conditions that pose challenges for accurate solar irradiance estimation.

\begin{table}[ht]
\centering
\resizebox{\textwidth}{!}{%
\begin{tabular}{c|c|cccccccccccccccccccc}
\toprule
\multicolumn{2}{c|}{\textbf{BSRN Stations}} & \textbf{ptr} & \textbf{daa} & \textbf{iza} & \textbf{pay} & \textbf{cnr} & \textbf{gob} & \textbf{cam} & \textbf{tam} & \textbf{ler} & \textbf{brb} & \textbf{flo} & \textbf{lin} & \textbf{tor} & \textbf{sms} & \textbf{cab} & \textbf{pal} & \textbf{ena} & \textbf{son} & \textbf{car} \\ 
\midrule

\multirow{3}{*}{\textbf{RMSE}} & \textbf{ERA5} & 61.40 & 44.85 & 87.68 & 59.01 & 58.42 & 42.91 & 47.16 & 53.55 & 52.08 & 66.25 & 74.68 & 40.37 & 40.05 & 66.00 & 41.78 & 46.01 & 60.55 & 66.64 & 42.61 \\ 
& \textbf{SARAH3 0.05$^\circ$} & 49.26 & 44.00 & 124.93 & 31.51 & 34.89 & 44.16 & 35.60 & 48.21 & 30.11 & 60.87 & 58.67 & 24.01 & 26.80 & 43.56 & 24.15 & 25.22 & 43.22 & 88.61 & 30.96 \\ 
& \textbf{SARAH3 0.25$^\circ$} & 45.14 & 43.50 & 112.81 & 29.84 & 33.00 & 43.90 & 31.21 & 48.29 & 29.08 & 57.83 & 54.21 & 23.02 & 25.29 & 43.46 & 23.73 & 24.84 & 38.50 & 72.07 & 29.64 \\ 
\midrule

\multirow{3}{*}{\textbf{MAE}} & \textbf{ERA5} & 31.45 & 20.32 & 55.51 & 30.94 & 29.61 & 16.61 & 24.52 & 28.19 & 24.45 & 34.63 & 38.96 & 21.00 & 19.66 & 33.71 & 21.42 & 22.97 & 31.18 & 35.41 & 20.33 \\ 
& \textbf{SARAH3 0.05$^\circ$} & 31.68 & 30.65 & 62.31 & 18.99 & 21.47 & 31.13 & 19.80 & 32.60 & 15.58 & 36.95 & 32.73 & 15.01 & 15.59 & 27.55 & 14.81 & 15.92 & 24.17 & 47.47 & 19.60 \\ 
& \textbf{SARAH3 0.25$^\circ$} & 29.64 & 30.37 & 60.68 & 18.39 & 20.73 & 30.98 & 18.06 & 32.76 & 15.21 & 35.71 & 30.58 & 14.45 & 14.95 & 27.58 & 14.61 & 15.69 & 22.13 & 39.33 & 19.17 \\ 
\midrule

\multirow{3}{*}{\textbf{Bias}} & \textbf{ERA5} & -3.19 & 1.30 & -43.91 & -0.67 & 2.30 & 3.07 & 1.43 & -7.12 & 13.09 & 3.84 & -0.33 & -2.64 & -1.24 & -6.91 & -2.09 & -2.62 & 10.42 & -10.25 & 0.70 \\ 
& \textbf{SARAH3 0.05$^\circ$} & 15.03 & -0.35 & -50.71 & 2.01 & 4.82 & 0.68 & -1.08 & -2.37 & -0.89 & 10.04 & 3.48 & 1.85 & -1.00 & -2.45 & 0.55 & 2.69 & 4.12 & -24.96 & 4.94 \\ 
& \textbf{SARAH3 0.25$^\circ$} & 11.26 & -0.83 & -52.04 & 1.78 & 3.35 & 1.39 & 1.63 & -1.79 & 0.08 & 8.51 & 3.36 & 0.10 & -1.34 & -3.65 & 0.38 & 1.94 & 3.29 & -17.10 & 4.94 \\ 
\midrule

\multirow{3}{*}{\textbf{Missing Values Ratio}} & \textbf{ERA5} & 0.08 & 0.20 & 0.02 & 0.01 & 0.01 & 0.01 & 0.34 & 0.02 & 0.08 & 0.56 & 0.06 & 0.00 & 0.00 & 0.07 & 0.02 & 0.02 & 0.01 & 0.10 & 0.07 \\ 
& \textbf{SARAH3 0.05$^\circ$} & 0.08 & 0.20 & 0.03 & 0.02 & 0.02 & 0.02 & 0.34 & 0.02 & 0.08 & 0.57 & 0.06 & 0.01 & 0.01 & 0.08 & 0.02 & 0.02 & 0.02 & 0.11 & 0.08 \\ 
& \textbf{SARAH3 0.25$^\circ$} & 0.08 & 0.20 & 0.03 & 0.02 & 0.02 & 0.02 & 0.34 & 0.02 & 0.08 & 0.57 & 0.06 & 0.01 & 0.01 & 0.08 & 0.02 & 0.02 & 0.02 & 0.11 & 0.08 \\ 

\midrule
\multicolumn{2}{c|}{\textbf{Latitude $^\circ N$}} & 67.57 & 75.43 & 28.30 & 46.82 & 45.20 & -23.00 & -23.85 & 22.79 & 27.65 & 13.16 & 51.93 & 60.12 & 43.56 & 34.40 & -34.91 & 38.00 & 50.52 & 32.25 & 69.67 \\ 
\midrule
\multicolumn{2}{c|}{\textbf{Longitude $^\circ E$}} & 24.09 & -25.45 & -16.50 & 6.94 & 5.28 & -47.62 & -46.07 & -3.45 & -15.58 & -59.45 & 4.93 & 10.42 & 1.33 & -6.88 & 20.81 & -1.48 & 7.61 & -110.95 & 20.93 \\ 
\midrule
\multicolumn{2}{c|}{\textbf{Altitude (m)}} & 20 & 11 & 2373 & 490 & 180 & 1050 & 878 & 11 & 100 & 4 & 100 & 50 & 800 & 800 & 150 & 700 & 100 & 200 & 20 \\ 
\bottomrule

\end{tabular}%
}
\caption{\textbf{Comparison of ERA5 and SARAH3 at BSRN locations.} ERA5 and SARAH3 6-hourly solar irradiance aggregations are compared against 6-hourly aggregations of BSRN observations in the SARAH3 domain. The comparison takes into consideration SARAH3 at original resolution (SARAH3 0.05$^\circ$) and the downsampled version (SARAH3 0.25$^\circ$). Except for the tam location, the downsampled SARAH3 dataset shows a lower RMSE. Compared to ERA5, SARAH3 demonstates higher accuracy for most of the locations.}

\label{tab:bsrn_era5_sarah3_comparison}
\end{table}

\begin{figure}[ht] 
    \centering 
    \includegraphics[width=\textwidth]{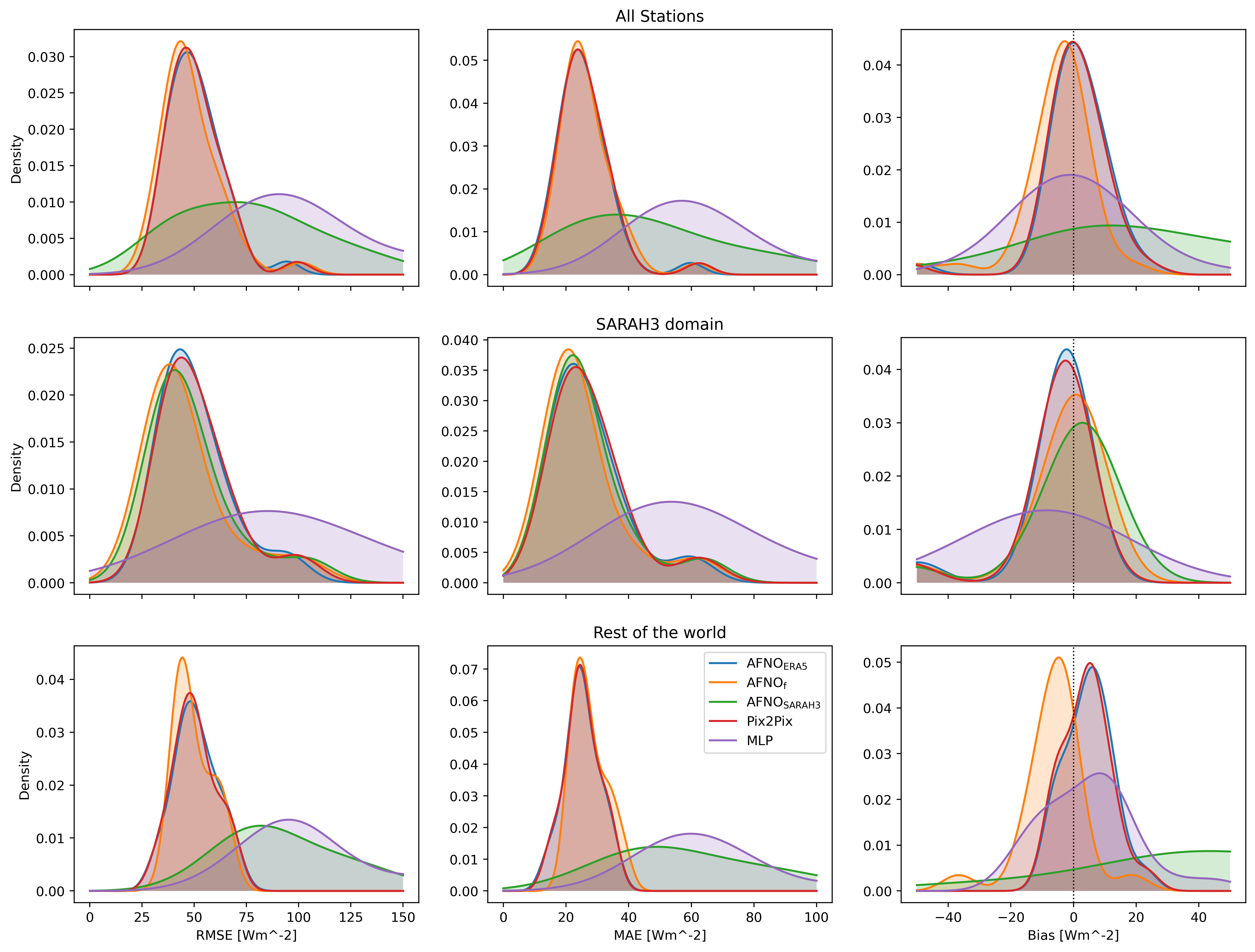} 
    \caption{\textbf{Evaluation vs BSRN 6-hour SSI}. RMSE, MAE and Bias are shown as density plots computed over three different subsets of BSRN stations for all the SSI estimation models. In the first raw, all the stations are used. In the second row, the predictions are compared to the stations locating in the SARAH3 domain. In the third row, the metrics are computed only on the stations outside the SARAH3 domain.} 
    \label{fig:density_plot} 
\end{figure}

\begin{figure}[h] 
    \centering 
    \includegraphics[width=\textwidth]{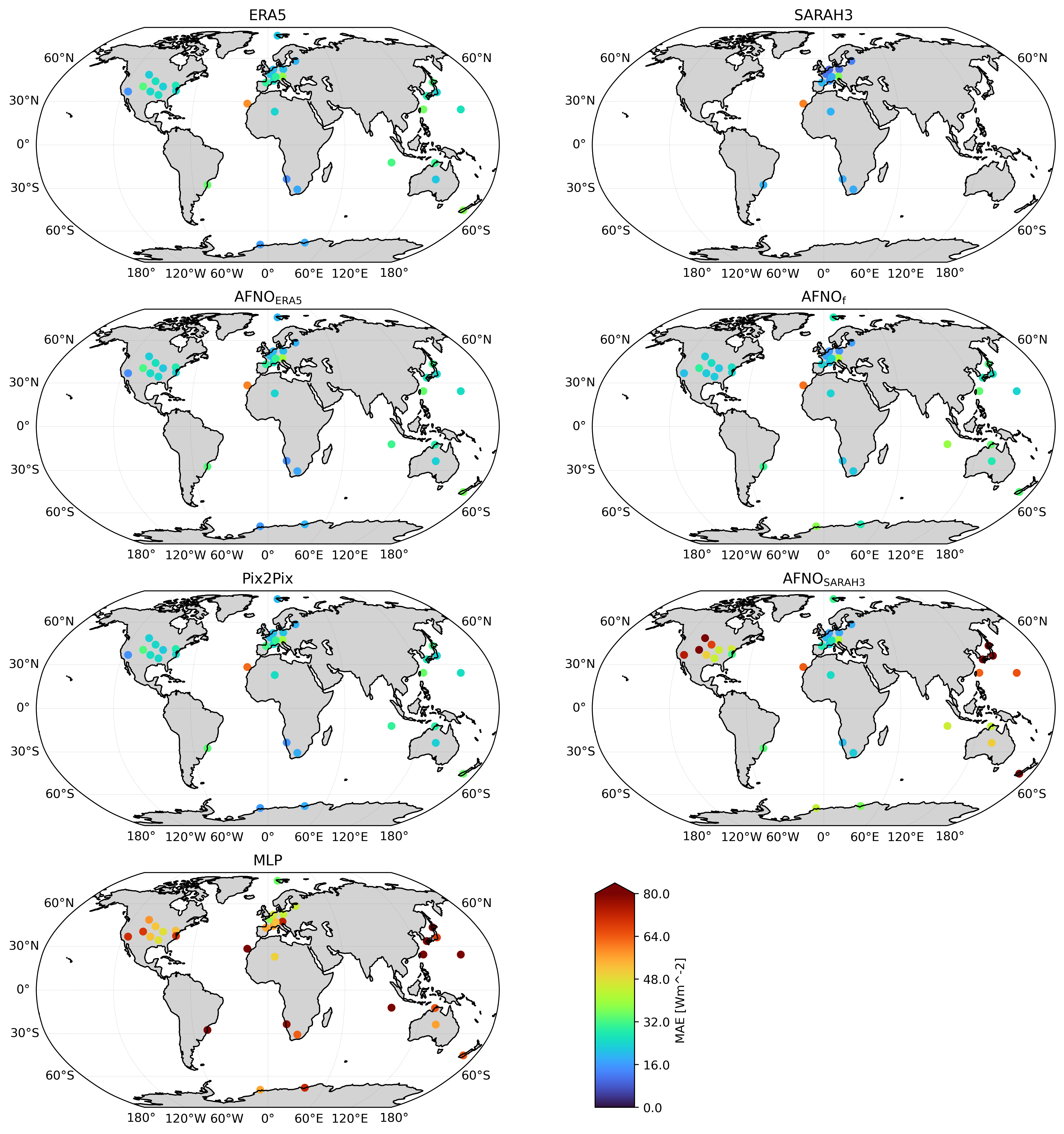} 
    \caption{\textbf{Mean Absolute Error} with respect to the the BSRN stations on 6h-aggregations for the test period (2018). On the left column, MAE is shown for ERA5 and models trained to replicate ERA5 SSI. On the right column, MAE is shown for SARAH3 and models trained on SARAH3 data.} 
    \label{fig:mae_gs} 
\end{figure}

\begin{figure}[h] 
    \centering 
    \includegraphics[width=\textwidth]{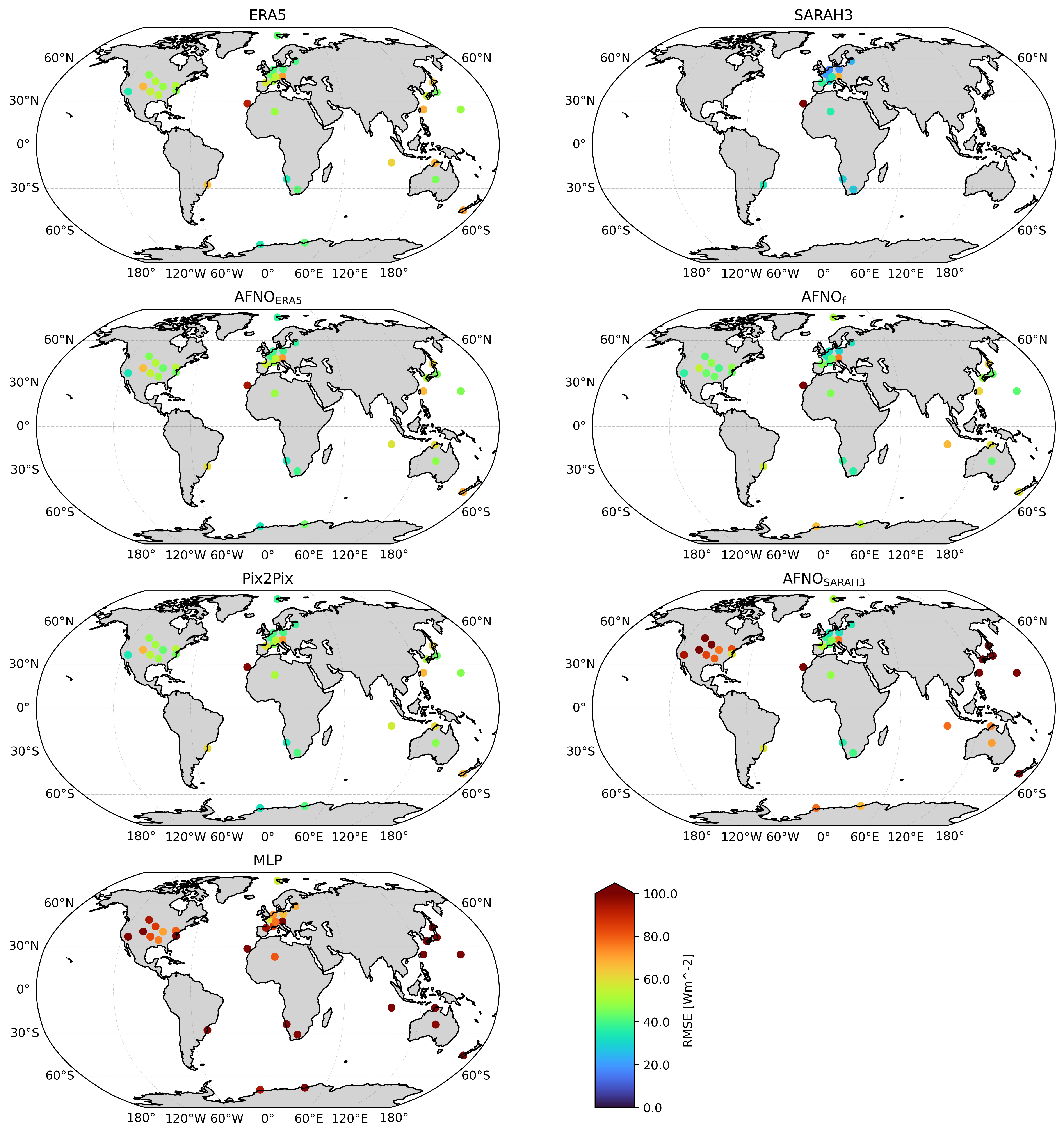} 
    \caption{\textbf{Root Mean Squared Error} with respect to the the BSRN stations on 6h-aggregations for the test period (2018). On the left column, RMSE is shown for ERA5 and models trained to replicate ERA5 SSI. On the right column, RMSE is shown for SARAH3 and models trained on SARAH3 data.} 
    \label{fig:rmse_gs} 
\end{figure}

\begin{figure}[h] 
    \centering 
    \includegraphics[width=\textwidth]{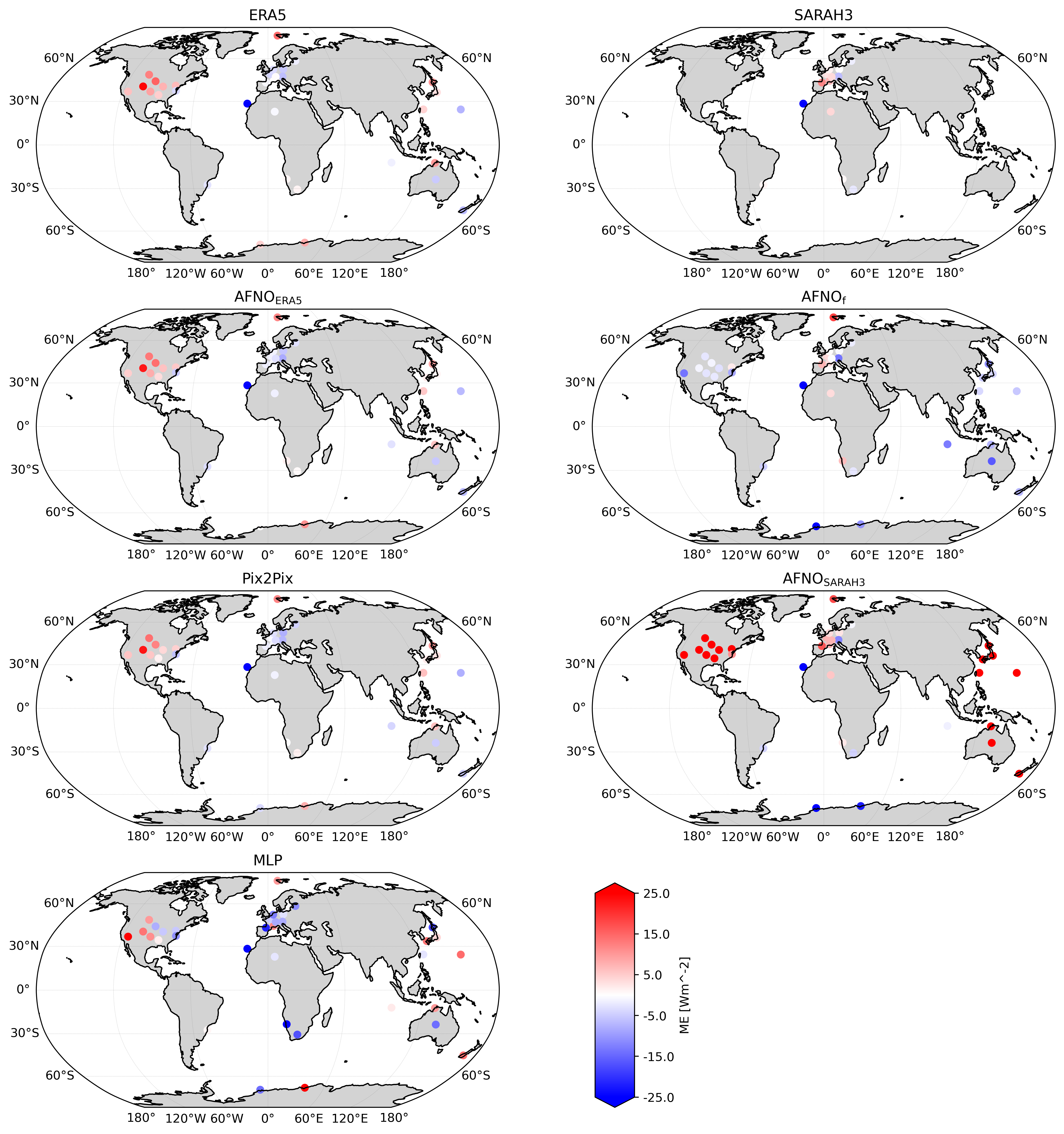} 
    \caption{\textbf{Mean error (bias)} with respect to the the BSRN stations on 6h-aggregations for the test period (2018). On the left column, ME is shown for ERA5 and models trained to replicate ERA5 SSI. On the right column, ME is shown for SARAH3 and models trained on SARAH3 data.} 
    \label{fig:bias_gs} 
\end{figure}

\clearpage

\end{document}